\documentclass{webofc}
\usepackage[varg]{txfonts}   % Web of Conferences font
\usepackage{hyperref}
\usepackage{url}
\usepackage{xcolor}
\usepackage{bm}
\usepackage{indentfirst}
\usepackage{subcaption}

\usepackage[normalem]{ulem}

\newcommand{\fzero}{$f_0(980)$}

\newcommand{\hcf}{Hydro-Coal-Frag}

\hypersetup{colorlinks=true,citecolor=blue,urlcolor=blue,linkcolor=blue}
\begin{document}
%
%\title{Probing the structure of \fzero\  with the elliptic flow in p-Pb collisions at LHC}
\title{Violation of the elliptic flow scaling of \fzero\  in p-Pb collisions at the LHC}

\author{
\firstname{Yili} \lastname{Wang}\inst{1}\fnsep\thanks{Speaker: Yili Wang, \email{coco.wyl@pku.edu.cn}, for Quark Matter 2025 proceedings}
\and
\firstname{Wenbin} \lastname{Zhao}\inst{2,3}
\and
\firstname{Che Ming} \lastname{Ko}\inst{4}
\and
\firstname{Feng-Kun} \lastname{Guo}\inst{5,6}
\and
\firstname{Ju-Jun} \lastname{Xie}\inst{7}
\and
\firstname{Huichao} \lastname{Song}\inst{1,8}
}

\institute{School of Physics, Peking University, Beijing 100871, China 
\and
Nuclear Science Division, Lawrence Berkeley National Laboratory, Berkeley, California 94720, USA
\and
Physics Department, University of California, Berkeley, California 94720, USA
\and
Cyclotron Institute and Department of Physics and Astronomy, Texas A\&M University, College Station, Texas 77843, USA
\and
Institute of Theoretical Physics, Chinese Academy of Sciences, Beijing 100190, China
\and
School of Physical Sciences, University of Chinese Academy of Sciences, Beijing 100049, China
\and
Institute of Modern Physics, Chinese Academy of Sciences, Lanzhou 730000, China
\and
Center for High Energy Physics, Peking University, Beijing 100871, China
}

\abstract{
We investigate the production and elliptic flow of the \fzero\ in high-multiplicity p–Pb collisions at $\sqrt{s_{NN}}=5.02$ TeV using a hadronic coalescence model with the $K$ and $\bar K$ phase-space distributions provided by the Hydro–Coal–Frag hybrid model. Our results, which agree with the ALICE and CMS measurements, support the $K\bar{K}$ molecular interpretation of the \fzero\ structure and show, however, a breakdown of the simple number-of-constituent (NC) scaling of its elliptic flow. The latter is in contrast to the deuteron elliptic flow, which exhibits a significantly better NC scaling when the same coalescence width parameter is used.
}
\maketitle
%
% \section{Introduction} 
% \label{sec:intro}

Relativistic heavy-ion and light-ion collisions at the Relativistic Heavy Ion Collider (RHIC) and the Large Hadron Collider (LHC) aim to create the quark–gluon plasma (QGP), a form of hot QCD matter that once existed in the very early universe. Near the critical temperature, the QGP undergoes hadronization, producing a variety of particles—including not only stable hadrons and resonances, but also light nuclei and exotic hadrons such as the deuteron, $X(3872)$, and $f_{0}(980)$. The formation of the QGP can lead to sufficiently large yields of certain light exotic hadrons, which enables systematic measurements of their transverse momentum spectra and anisotropic flows, offering a unique opportunity to probe the internal structure of exotic hadrons, e.g., to distinguish between compact tetraquark states and loosely bound molecular configurations. The $f_{0}(980)$, reconstructed via the decay channel $f_{0}(980) \to \pi^+ \pi^-$ [B.R. = (46 ± 6)\%], is an especially promising candidate due to its relatively low mass and abundant production. Recently, its yield, transverse momentum spectra, and elliptic flow ($v_2$) have been measured in p–Pb collisions at the LHC~\cite{collaborationObservationAbnormalSuppression2024,
collaborationEllipticAnisotropyMeasurement2023}. Analysis of the $v_2$ data from the CMS Collaboration, using the number-of-constituent-quark (NCQ) scaling, suggests that the $f_0(980)$ has a conventional $q\bar{q}$ structure. This conclusion is in contrast with the hadron physics studies, which provide substantial evidence for a tetraquark or $K\bar{K}$ molecular configuration~\cite{Jaffe:1976ig,Jaffe:1976ih,weinsteinMolecules1990,Baru:2003qq, guoHadronicMolecules2018}. In this work, we employ a hadronic coalescence model, which is commonly used to study the production of light nuclei, together with phase-space distributions of kaons from the Hydro–Coal–Frag hybrid model to calculate the production and elliptic flow of the $f_{0}(980)$ in p–Pb collisions at $\sqrt{s_{NN}} = 5.02$  TeV. Our results support a molecular interpretation of the \fzero\ and indicate that the hadronic coalescence process leads to a deviation from the naive NC scaling, which is more pronounced for the \fzero\ than for the deuteron. This proceedings contribution is based on our work in Ref.~\cite{Wang:2025prep}.

% \section{Methodology}  \label{sec:meth}

\begin{figure*}[ht]
    \centering
    \includegraphics[width=1.0\linewidth]{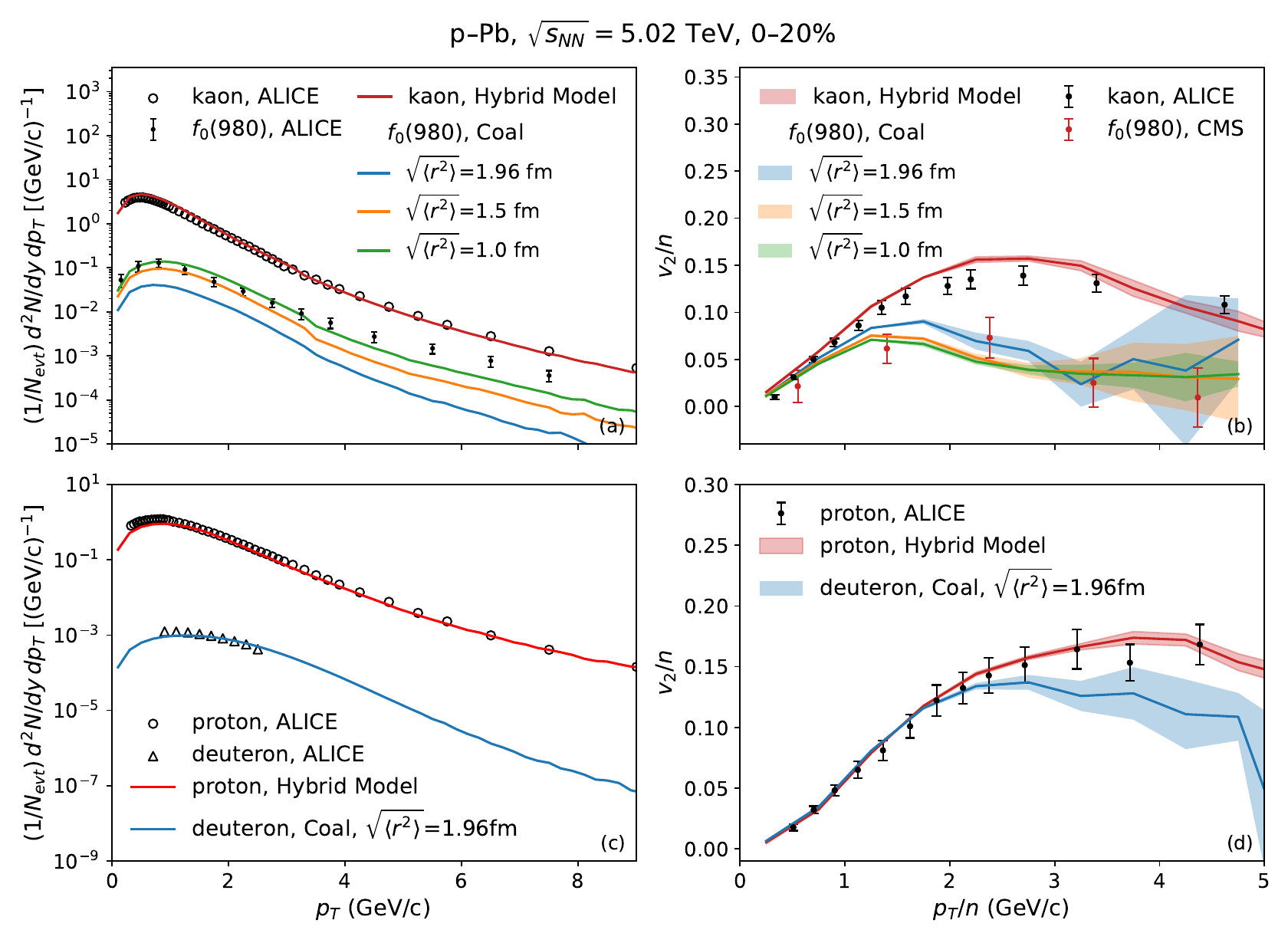}
    \caption{$p_T$-spectra and $v_2(p_T)$ of \fzero\  and  deuterons calculated from the coalescence model, using the phase-space distributions  of kaons and  protons  from the \hcf\ Hybrid model~\cite{Zhao:2020wcd,YuanyuanWang:2023wbz}. Data are from~\cite{collaborationMultiplicityDependenceCharged2016,collaborationObservationAbnormalSuppression2024,ALICE:2024vzv,collaborationEllipticAnisotropyMeasurement2023,ALICE:2019bnp}.}
    \label{fig:spectra_v2}
\end{figure*}

The $f_{0}(980)$ is a resonance with a finite width near the $K\bar{K}$ threshold. Its $K\bar{K}$ component exhibits characteristic features of a weakly bound system, similar to the deuteron~\cite{guoHadronicMolecules2018}. We therefore assume that it survives only after kinetic freeze-out and calculate its yield using the hadronic coalescence approach. In this model, the formation probability of a weakly bound composite is determined by the overlap of the Wigner function, $\rho^{W}$, with the phase-space distributions of its constituent particles, $f_i(\mathbf{x}_i, \mathbf{p}_i, t_i)$, at the kinetic freeze-out \cite{chenLightClusterProduction2003,Zhao:2018lyf,Zhao:2020irc}:
%\fk{Shouldn't $\frac{dN}{d^3\bm{P}}$ be $\frac{d^3N}{d\bm{P}^3}$?}
\begin{align}
     \frac{dN}{d^3\bm{P}} = g \int p_1^\mu d^3\sigma_{1\mu}\frac{d^3\bm{p}_1}{E_1}\, p_2^\nu d^3\sigma_{2\nu}\frac{d^3\bm{p}_2}{E_2}\notag& 
\times  f_{1}(\bm{x}_1, \bm{p}_1,t_1) \, f_{2}(\bm{x}_2, \bm{p}_2,t_2) \notag \\
&\times \rho^{W}(\bm{x}'_1, \bm{x}'_2;\bm{p}'_1,\bm{p}'_2;t')\, \delta(\bm{P} - \bm{p}_1 - \bm{p}_2),
\end{align}
where $g = (2J + 1) / \prod_{i=1}^A (2J_i + 1)$ is the statistical factor for forming a composite with angular momentum $J$ from $A$ constituents of spin $J_i$.  In the above, $\bm{x}_i$ and $\bm{p}_i$ are the coordinates and momenta in the lab frame, while $\bm{x}'_i$ and $\bm{p}'_i$ are defined in the \fzero\ rest frame. Assuming a Gaussian wave function for the formed composite, its Wigner function is then given by
\begin{align}
    \rho^{W}(\boldsymbol\rho,{\bf p}_\rho)=8\exp\left[-\frac{\boldsymbol\rho^2}{\sigma_{\rho}^2}-{\bf p}_\rho^2\sigma_{\rho}^2\right], 
    \label{rho}
\end{align}
where $\boldsymbol{\rho} = \frac{1}{\sqrt{2}}(\mathbf{x}'_1 - \mathbf{x}'_2)$ and $\bf p_\rho=\frac{1}{\sqrt{2}}({\bf p}_1^\prime-{\bf p}_2^\prime)$ are the relative coordinate and momentum in the composite rest frame. The Gaussian width $\sigma_{\rho}$ is related to the root-mean-square (RMS) radius $\sqrt{\langle r^2 \rangle}$ of the composite via $\sigma_{\rho} = \frac{2}{\sqrt{3}} \sqrt{\langle r^2 \rangle}.$ For the deuteron, experimental data gives an RMS radius $\sqrt{\langle r^2 \rangle}=$1.96~fm~\cite{HypHI:2013sxa}, while the RMS radius of the $f_{0}(980)$ remains uncertain due to limited data. Following the standard estimate for near-threshold hadronic molecules \cite{guoHadronicMolecules2018}, its size can be approximated as $R \sim  1/ \sqrt{2\mu E_B} $, where $\mu$ is the reduced mass and $E_B$ the binding energy. For $E_B \approx 10\text{–}20$~MeV, the RMS radius of the $f_{0}(980)$ is estimated to be $\sqrt{\langle r^2 \rangle}\sim 1.0\text{–}1.5$ fm. In the following calculations in Fig.~\ref{fig:spectra_v2}, we also increase the RMS radius of the $f_{0}(980)$ to $\sqrt{\langle r^2 \rangle}=$1.96~fm, just for a comparison with the calculations of the deuteron. The phase-space distributions $f_i(\bm{x}_i,\bm{p}_i,t_i)$ are taken from the \hcf\ hybrid  model, which combines hadron production at low $p_T$ from hydrodynamics, at intermediate $p_T$ from quark coalescence, and at high $p_T$ from string fragmentation~\cite{Zhao:2020wcd}. This model has been tuned to reproduce the $p_T$ spectra and the elliptic flow of pions, kaons, and protons up to 6 GeV in high multiplicity p–Pb collisions at the LHC~\cite{YuanyuanWang:2023wbz}, relevant for the production of $f_{0}(980)$ and deuterons via $K\bar{K}$ and $pn$ coalescences. 
% \section{Results and Discussion}  \label{sec:res}

%and $p_T$-spectra and elliptic flow ($v_2$) of kaons (upper panel) and  protons (lower panel) from the \hcf\ model~\cite{Zhao:2020wcd}.

%$p_T$-spectra and elliptic flow ($v_2$) of kaons 

Upper panels of Fig.~\ref{fig:spectra_v2} show the $p_T$ spectra and $v_2(p_T)$ of kaons and $f_{0}(980)$  in p-Pb collisions at $\sqrt{s_{NN}}=5.02$~TeV. The results of kaons are calculated from the \hcf\ model, which well reproduce the ALICE and CMS data up to 6 GeV. Using the $K$ and $\bar{K}$ phase-space distributions at the kinetic freeze-out, we compute the $f_{0}(980)$ production via $K+\bar{K}\to f_{0}(980)$ coalescence. The resulting spectra and $v_2$ are consistent with the ALICE and CMS measurements, with $\sqrt{\langle r^2 \rangle}=$ 1.0 fm or 1.5 fm. Although the RMS radius has a noticeable effect on the yields of the $f_{0}(980)$, it only slightly influences its $v_2(p_T)$. A mild underestimation of the yields at high $p_T$ is likely due to the absence of fragmentation contributions. Overall, the good agreement with the $v_2$ data supports the $K\bar{K}$ molecular interpretation of the $f_{0}(980)$. The upper right panel demonstrates, however, a breakdown of the NC scaling of $v_2$ for the $f_{0}(980)$ because its Wigner function  depends on both coordinate and momentum separations.  A small $\sigma_\rho$ (with $\sqrt{\langle r^2 \rangle}=$ 1.0 fm or 1.5 fm) broadens the momentum width, allowing  recombination of constituents with unequal momenta and leading to the violation of the NC scaling of $v_2$~\cite{Wang:2025prep}.

For comparison, we also compute deuteron production via $p + n \to d$ coalescence with the coalescence parameter $\sigma_\rho=2.26$ fm ($\sqrt{\langle r^2 \rangle}=$ 1.96 fm). As shown in lower panels of Fig.~\ref{fig:spectra_v2}, the $p_T$-spectra and $v_2(p_T)$ of protons are well reproduced, and the deuteron spectra from the coalescence model fit the measured data well. The upper and lower right panels also show that, with the same RMS radius $\sqrt{\langle r^2 \rangle}=$ 1.96 fm, a substantially better NC scaling of $v_2$ for the deuteron than for the $f_{0}(980)$. This difference can be attributed to the much higher density of kaons than that of protons and neutrons at the kinetic freeze-out, which increases the coalescence probability of  kaons with mismatched spatial coordinates and momenta associated with the NC scaling violation~\cite{Wang:2025prep}.

%Moreover, the deuteron contains 6 valence quarks, whereas the $f_0(980)$, being a flavor-neutral meson, can be produced from a quark-antiquark pair even if it is predominantly a $K\bar{K}$ molecular state.  %\fk{This comment is included for your reference and may be removed later.} \com{\bf[Removing the last setence would reduce the paper to the required 4 pages.]}

In summary,  we have calculated the $p_T$-spectra and elliptic flow $v_2(p_T)$ of the $f_{0}(980)$ in high-multiplicity p–Pb collisions at $\sqrt{s_{NN}} = 5.02$ TeV, using a hadronic coalescence model with $K+\bar{K}\to f_{0}(980)$ coalescence. Our calculations reproduce the measured data well, which supports the \fzero\ being a loosely bound $K\bar{K}$ molecule. 
We further show that, with reasonable values for the RMS radius of $f_{0}(980)$,  
the Wigner function in the $K\bar{K}$ coalescence allows recombination of kaons with unequal momenta, leading to a breakdown of the hadronic NC scaling of $v_2$ for the $f_{0}(980)$. In contrast, the deuteron exhibits a significantly better NC scaling behavior of $v_2$, due to the lower densities of its constituent protons and neutrons at the kinetic freeze-out than those of kaons. \\

\noindent \textsl{Acknowledgments.} Y. W. and H. S. are supported by NSFC under Grants No. 12575138 and No. 12247107. F. G. and J. X. are supported by NSFC under Grants No. 12125507, No. 12361141819, No. 12447101, and No. 12435007. W. Z. is supported by the NSF under grant number ACI-2004571 within the JETSCAPE collaboration and by the DOE within the SURGE Collaboration. C. M. K. is supported by the DOE under Award No. DE-SC0015266. 

%
% BibTeX or Biber users please use (the style is already called in the class, ensure that the "woc.bst" style is in your local directory)
% \bibliography{your_bib_file} % Replace "your_bib_file" with the actual name of your .bib file
% \bibliographystyle{elsarticle-num}  

% \bibliographystyle{unsrt}
\bibliography{references}

\end{document}